\documentclass[%
 reprint,
 prl,
 amsmath,amssymb,
 aps,
 longbibliography
]{revtex4-1}
\usepackage[utf8]{inputenc}
\usepackage{amsmath,amssymb}
\usepackage{bm}
\usepackage{graphicx}
\usepackage{float}
\usepackage{amsthm}
\usepackage{braket}
\usepackage{mathtools}
\usepackage[colorlinks,linkcolor=blue,citecolor=blue]{hyperref}
\usepackage{version}	% required for `\comment' (yatex added)
\begin{document}
\title{Capture and chaotic scattering of a charged particle by magnetic monopole
under uniform electric field}

\author{Kou Misaki,${}^{1}$ Naoto Nagaosa${}^{1,2}$}

\affiliation{
$^1$Department of Applied Physics, The University of Tokyo, 
Bunkyo, Tokyo 113-8656, Japan
\\
$^2$RIKEN Center for Emergent Matter Science (CEMS), Wako, Saitama 351-0198, Japan
\\
}
\date{\today}
\begin{abstract}
Motivated by the realization of magnetic monopole of Berry curvature
by the energy crossing point, we theoretically study
the effect of magnetic monopole under a uniform electric field
in the semiclassical dynamics, which is relevant to many physical
 situations such as relaxation through the diabolic point.
We found that the competition between the backward scattering by
 the monopole magnetic field and the acceleration by the electric field
 leads to the bound state, i.e., capture of a particle near the
 monopole. Furthermore, the nonlinearity induced
by the magnetic monopole leads to the
chaotic behavior in the transient dynamics, i.e., the transient
chaos. We computed characteristic quantities of the strange
saddle which gives rise to the transient chaos, and verified that the
abrupt bifurcation occurs as we tune the system parameter toward the
parameter region in which the system is solvable.
\end{abstract}
\maketitle

{\it Introduction.---}
Since Dirac theoretically pointed out its possible existence by considering the
$2\pi$ ambiguity of the phase of an electron wavefunction
\cite{dirac1931quantised}, magnetic monopole
has attracted wide theoretical interests and
appeared in many areas of physics
\cite{Hooft1974,polyakov1974,castelnovo2008magnetic}.
In particular, magnetic monopole
of Berry curvature \cite{berry1984quantal} or synthetic
gauge field \cite{dalibard2011colloquium,goldman2014light} appears both
in real \cite{Zhang2005,Ruseckas2005,ray2014observation}
and momentum
\cite{murakami2007phase,wan2011topological,hosur2013recent,armitage2017weyl}
spaces, and drastically affects the transport
property as it modifies the semiclassical equation of motion for the wave
packet of particles
\cite{berry1993classical,berry1993chaotic,sundaram1999wave,xiao2010berry}.

Actually, the history of magnetic monopole
\cite{goddard1978magnetic,preskill1984magnetic,tong2005,Shnir2005,mcdonaldbirkeland}
dates back to the late 1800s when Darboux and Poincar\'e
theoretically studied the
scattering problem of an electron by a magnetic monopole
\cite{Darboux1878,poincare1896remarques}.
This problem can be exactly
solved because of the conservation of the angular momentum, and exhibits
some unusual properties compared to the usual potential scattering
\cite{Shnir2005,Note1}. In this paper, we will show that, upon
introducing the uniform electric field, the peculiar
nature of this scattering problem leads to the chaotic dynamics, i.e.,
the chaotic scattering \cite{Gaspard1989,Bleher1990}.

\begin{figure}
 \centering
 \includegraphics[width=\hsize]{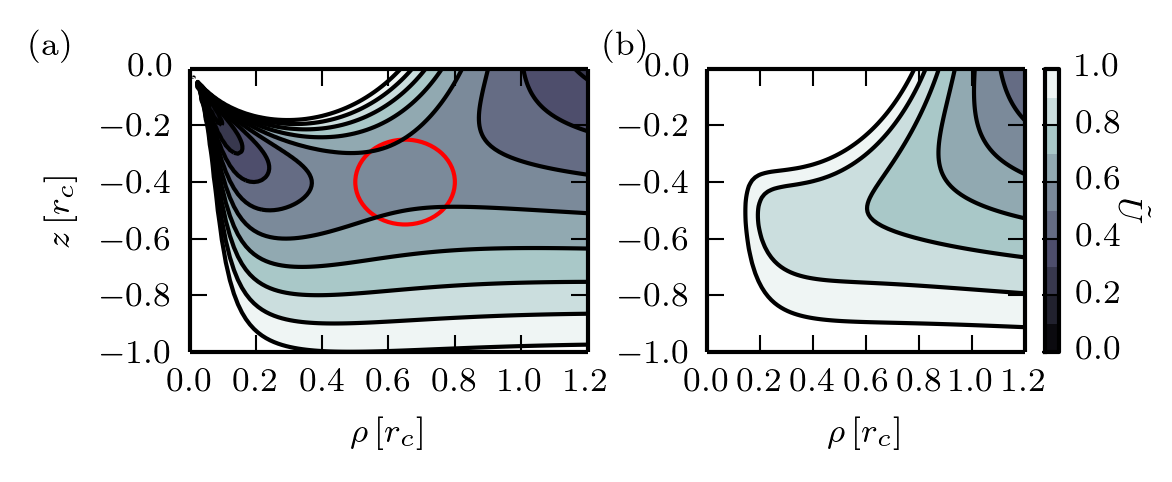}
 \caption{The distribution of $U_\text{eff}$ in $(\rho,z)$ plane
 for (a) $J_z/(q_mq_e)=0.9$, and (b)
 $J_z/(q_mq_e)=\sqrt{32/27}+0.01$.
 The color represents the potential height
 $\tilde{U}\coloneqq U_\text{eff}/(fr_c)$, where
 $r_c\coloneqq |q_mq_e|^{2/3}/(mf)^{1/3}$, as is shown in the right of
 (b).
 We can see the potential saddle as we highlighted with the red circle
 for (a), while the saddle disappears for (b).}
  \label{potential}
\end{figure}

{\it Model.---}
We numerically study the equation describing the dynamics
of a particle under the influence of monopole magnetic field and the
uniform electric field:
\begin{equation}
 m\frac{d^2\vec{r}}{dt^2}=f
 \vec{e}_z+q_m q_e\frac{d\vec{r}}{dt}\times
 \frac{\vec{r}}{r^3}, \label{EoM}
\end{equation}
where $m$ is the mass of particle, $f$ is the uniform force along $z$
direction, 
$q_m$ and $q_e$ are the magnetic charge of the monopole sitting at the
origin and the electric
charge of the particle, respectively.
This equation has two conserved quantities, i.e., the energy and the angular
momentum along $z$ direction:
\begin{equation}
 E=\frac{m}{2}(\dot{\vec{r}})^2-fz,\quad
  J_z=m(x\dot{y}-y\dot{x})-q_m q_e\frac{z}{r}, \label{EandJz}
\end{equation}
where the second term of $J_z$ comes from the angular momentum of the
electromagnetic field. We can utilize $J_z$ conservation to
reduce one degree of freedom \footnote{See supplemental material
for the derivation of Eq. (\ref{simpEoM}), the details of the form of
$U_\text{eff}$ and capturing, the origin of the peak structure of
$t_\text{pass}$,
the details of the exact
solution of Eq. (\ref{EoM}) with $f=0$, the calculation of the Lyapunov
exponent and the escape rate, and the effect of the mass deformation and
dissipation.}:
\begin{align}
  &\left\{
 \begin{alignedat}{2}
  m\frac{d^2 z}{dt^2}&=-\frac{\partial U_\text{eff}}{\partial z}\\
  m\frac{d^2 \rho}{dt^2}&=-\frac{\partial U_\text{eff}}{\partial \rho},
  \end{alignedat}
 \right.
 \quad
U_\text{eff}\coloneqq -fz+\frac{\left(J_z+ \frac{q_m q_e
 z}{\sqrt{z^2+\rho^2}}\right)^2}{2m\rho^2}, 
 \label{simpEoM}
\end{align}
where $\rho=\sqrt{x^2+y^2}$.
Now the system is fully characterized by $U_\text{eff}$, and the problem
reduces to the usual potential scattering. The form of the potential
depends on the value of $J_z$, which is determined from the initial
conditions according to Eq. (\ref{EandJz}). As we can see from the
Fig. \ref{potential} (a), there exists a potential saddle
for particular parameter range of $J_z$.
If we consider the dynamics of the particle starting from the initial
position inside the potential pocket bounded by the saddle and
the potential walls, the particle bounces
back and forth inside the pocket, and
eventually goes over the saddle.
Therefore, the existence of the
saddle and the potential pocket it bounds
are the crucial ingredients for this scattering problem.
It turns out that the saddle exists for $|J_z|<\sqrt{32/27}|q_mq_e|$
\cite{Note1}, so we concentrate on the dynamics for this range of $J_z$.

Since the diverging magnetic field leads to the infinite cyclotron
frequency, it invalidates the numerical calculation for the trajectory
passing near the origin. To avoid this difficulty, we introduce the
smeared magnetic charge:
\begin{equation}
\rho(\vec{r})=\frac{1}{\xi^3\sqrt{\pi^3}}\int d\vec{R}\,
 e^{-\frac{|\vec{r}-\vec{R}|^2}{\xi^2}}q_m\delta(\vec{R})
 =\frac{q_m e^{-\frac{r^2}{\xi^2}}}{\xi^3\sqrt{\pi^3}}. \label{smearmag}
\end{equation}
The magnetic field produced by this magnetic charge is the same as the
monopole for $r \gg \xi$, and converges to $0$ as $r\to 0$.

\begin{figure}
 \centering
 \includegraphics[width=\hsize]{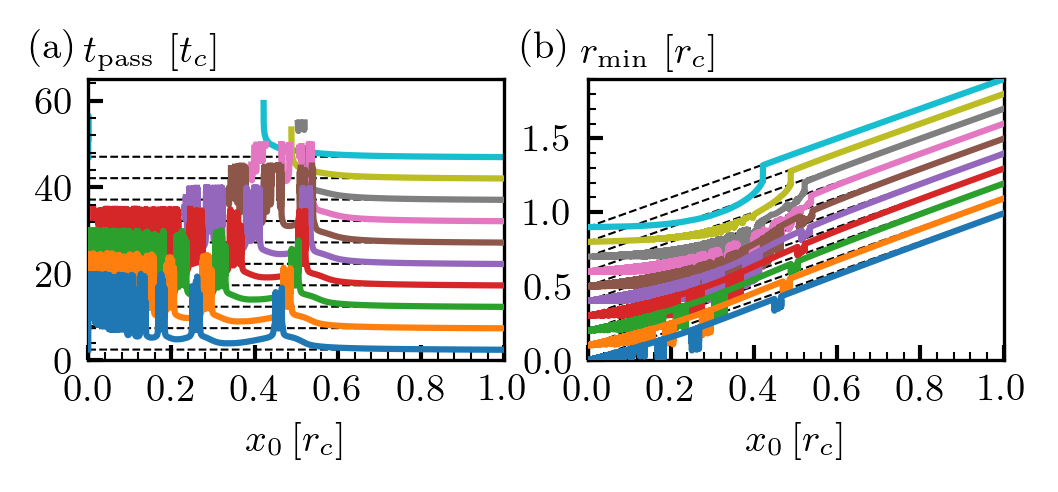}
 \includegraphics[width=\hsize]{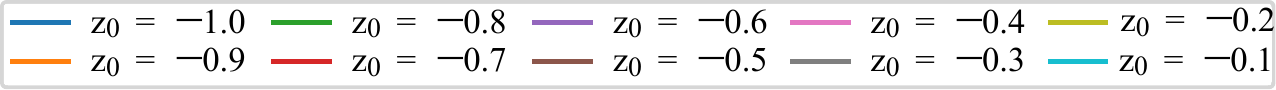}
 \caption{The plot of the time when the particle get out of the
 scattering region $r \leq 2r_c$
 ($t_\text{pass}$) and the minimum value of $r(t)$
 ($r_\text{min}\coloneqq \min_{t}(r(t))$), for
 the initial position $(x_0,0,z_0)$. Here $r_c\coloneqq
 |q_mq_e|^{2/3}/(mf)^{1/3} $ and $t_c\coloneqq (m
 |q_mq_e|)^{1/3}/f^{2/3}$. The numerical calculation was done with an
 implicit Tajima method. The black dashed line represent the values
 without the monopole magnetic field. Note that each plot is shown with
 an offset. The values of the offset are given by the values of the
 dashed line at $x_0=0$. For small $x_0$ and $z_0$, $t_\text{pass}$ is
 larger than $t/t_c=20$, and the plot of $t_\text{pass}$ at these values
 are not shown.}
  \label{passtrmin}
\end{figure}

\begin{figure}
 \centering
 \includegraphics{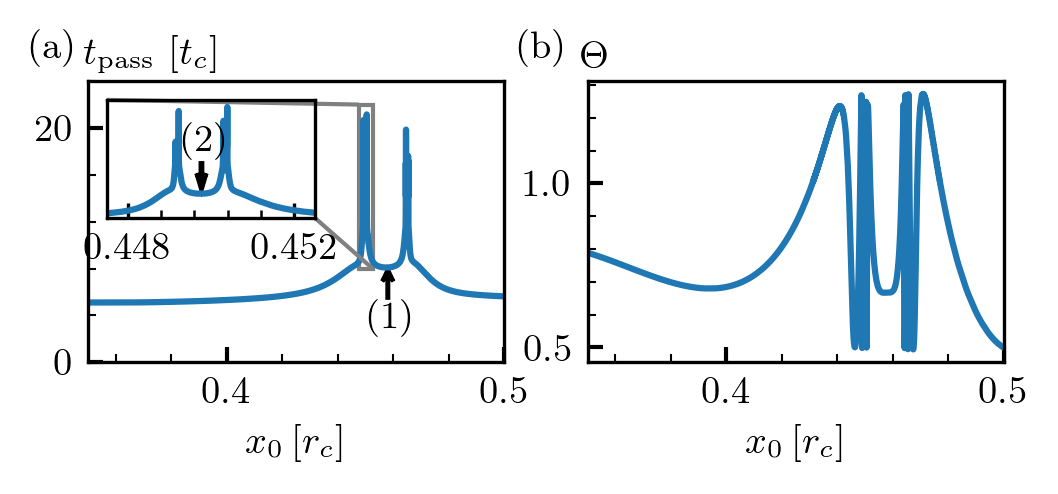}
 \caption{(a) The peak structure of $t_\text{pass}$
 around $x_0/r_{c} \cong 0.45$ at
 $z_0/r_{c}=-1$. The inset shows the detailed peak structure of the left
 peak, and we can clearly see the self-similarity of the peak.
 We showed $(1)$ and $(2)$, which is the parameter point
 we take for the dynamics shown in Fig. \ref{frac}. We define the
 parameter points $(n)$ for all $n$ larger than $2$ for finer
 structures in the same way.
 (b) The rapid
 variation of $\Theta \coloneqq
 \cos^{-1}(v_z(t_\text{pass})/\sqrt{\vec{v}(t_\text{pass})\cdot\vec{v}(t_\text{pass})})$
 near the fractal peak, which is the characteristic feature of the
 chaotic scattering \cite{Gaspard1989,Bleher1990}.}
  \label{frachie}
\end{figure}

\begin{figure}
 \centering
 \includegraphics[width=\hsize]{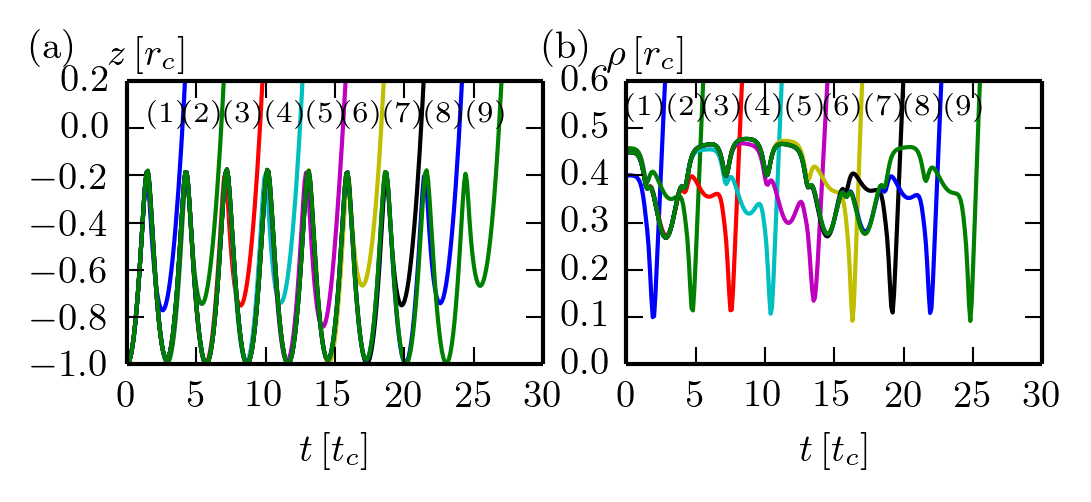}
 \caption{The plots of (a) $z(t)$ and (b) $\rho(t)$.
 The value of $x_0$ at $(n)$ is defined in Fig. \ref{frachie}. We can
 see the correspondence between the number of the oscillation and the
 number of the hierarchy.}
  \label{frac}
\end{figure}

{\it Result.---}
We performed the numerical calculation of Eq. (\ref{EoM}) with the
Runge-Kutta method and the implicit Tajima method \cite{1991plasma}.
We can regard the dynamics described by Eq. (\ref{EoM}) as
a scattering problem of a charged particle by magnetic monopole. There are two
important physical observables: $t_{\text{pass}}$, which is the time it
takes for the particle to get out of the scattering region, which we
define as $r \leq 2r_c$ ($r_c\coloneqq |q_mq_e|^{2/3}/(mf)^{1/3}$), and
$r_{\text{min}}\coloneqq \min_{t}\{r(t)\}$, which determines whether our
approximation of the point magnetic charge by the smeared one is good or
not. We set the initial velocity to be zero, and vary
$z(0)=z_0$ and $x(0)=x_0$ to adjust the incident velocity and the impact
parameter, respectively, and set $q_mq_e<0$.
We show the result of the numerical calculation
in Fig. \ref{passtrmin}. For small $x_0$ (small $J_z$) and small $|z_0|$ (small
energy) region, we observed that $t_\text{pass}$ becomes larger than the
numerically accessible time region.
This fact can be understood from the
geometry of the effective potential Eq. (\ref{simpEoM}): For this
region, the particle cannot escape from the scattering region
since the height of the potential saddle is higher than the initial energy.
We obtained the analytic
expression for the parameter region where this trapping occurs,
and verified that the expression matches
with the result of the numerical calculation \cite{Note1}.
In addition, we can see the complicated peak structure in
the region with small impact parameter, and as is shown in
Fig. \ref{frachie}(a),
each peak has a fractal structure, and the scattering angle $\Theta$,
which is the relative angle between the initial and the final velocity, varies
very wildly near this peak structure.
We show the dynamics at each
hierarchy of the fractal in Fig. \ref{frac}.
We checked the convergence of the peak
structure and the small error of conserved quantities, so
this peak structure is not an
artifact of the finite precision of the numerical calculation.
Actually, this fractal structure of $t_\text{pass}$ is a characteristic
feature of the chaotic scattering \cite{Gaspard1989,Bleher1990}, which
is the representative example of the
transient chaos \cite{ott2002,lai2011}.

To understand why the chaotic scattering occurs and the multiple peaks
for each $z_0$ is observed, it is convenient to go back to $f=0$ limit
of Eq. (\ref{EoM}). As we mentioned in the introduction, this model is
exactly solved.
Among some peculiar
characteristics, one notable feature is
the impact parameter ($x_0$) dependence of the scattering angle: There are
multiple backscattering points (so-called glory scattering) located at
$q_e q_m/(mvx_0)=\sqrt{4n^2-1}$ $(n=1,2,\dots)$, where $v$
represents the initial velocity
\cite{boulware1976scattering,SCHWINGER1976451}.
This is in sharp
contrast to the scattering angle
in the Rutherford scattering which has a monotonic impact
parameter dependence. If we regard each backscattering point as the
potential hill, the situation is formally similar to the scattering
problem by multiple scatterer, which is known to exhibit the chaotic
behavior at the transient time scale \cite{Gaspard1989,Bleher1990}.
Moreover, for fixed $z_0$ there are infinitely many $x_0$ where the
particle is backscattered, and this leads to infinitely many well-separated
peak structures \cite{Note1}.

The chaotic scattering
can be understood as a consequence of the horseshoe type
mapping induced by a intersection of a stable and
an unstable manifold of the periodic orbits
\cite{smale1967differentiable}. This mapping leads to the invariant
saddle which has a fractional dimension.
The fact
that the saddle has a fractal dimension has a drastic consequence on the
physical quantities: The dimension of generic crossing of the stable
manifold of the saddle and the one parameter family of initial
conditions in the phase space becomes fractional.
As a result, at each intersection, the
time it takes for the particle to get out of the scattering region
($t_\text{pass}$) is infinite, so $t_\text{pass}$ has a peak on the set
with a fractional dimension. The dimension of the crossing can be
calculated as follows \cite{CHEN199093}:
Our model is the
Hamiltonian system with two degrees of freedom (Eq. (\ref{simpEoM})).
Within each energy shell,
the dimension of the phase space is three. Noting that
Eq. (\ref{simpEoM}) has time reversal symmetry, the dimensions of the
stable and the unstable manifold are the same.
Since the dimension of the intersection of the two
subsets $S_1$ and $S_2$ in the $d$ dimensional manifold is given by
$D(S_1 \cap S_2)=D(S_1)+D(S_2)-d$, where $D(S)$ represents the dimension
of $S$, and the saddle is given by the intersection of the stable and the
unstable manifold, $D(S_\text{st})=(D(S_\text{sad})+3)/2$, where
$S_\text{sad}$ and $S_\text{st}$ represent the saddle and the stable
manifold of the saddle. Therefore, the dimension of the crossing, which
we refer to as the fractal dimension, is
given by $d_\text{fra}=1+D(S_\text{st})-3=(D(S_\text{sad})-1)/2$.

Concerning the dynamics on the saddle,
the positivity of the Lyapunov exponent itself does not immediately
imply the chaotic behavior, which can be quantitatively understood from
the following Kantz-Grassberger formula
\cite{KANTZ198575,Bleher1990,gaspard2005chaos,lai2011}:
$\sum_{\lambda_i>0}\lambda_i=\kappa+h_\text{KS}$, where $\lambda_i$s are
the Lyapunov exponents of the invariant set, $\kappa$ is the escape rate, i.e.,
$\kappa\coloneqq\lim_{T\to\infty}-(1/T)\ln(N(T)/N(0))$ ($N(T)$ is the
number of particles
remaining in the scattering region at $t=T$), and $h_\text{KS}$ is the
Kolmogorov-Sinai entropy, of which nonzero value implies the
chaos.
This equation represents the fact that the instability of
the invariant set (Lyapunov exponent) leads to two phenomena: The
escape of the particle from the scattering region and the growth of the
information which is the characteristic feature of the chaos.
As we can see, positive $\lambda_i$ with $h_\text{KS}=0$ is possible
because of the finite escape rate $\kappa$ in contrast to the case with
the attractor.
For the
numerical calculation, the following Young's formula
\cite{young1982dimension,gaspard2005chaos} is useful: For the
Hamiltonian system with two degrees of freedom, it is
$h_\text{KS}=\lambda(D_1-1)/2$, where $D_1$ is the
information dimension and is defined by
\begin{equation}
 D_1\coloneqq \lim_{q\to 1}D_q\coloneqq \lim_{q\to 1}\lim_{\delta\to
  0}\frac{1}{q-1}\frac{\log\sum_ip_i^q}{\log\delta}, \label{Renyi}
\end{equation}
where $\delta$ is the linear size of the box we use to divide the
phase space to define the measure $p_i$, and $D_q$ is the R\'enyi
dimension. Compared to the box counting dimension of the invariant set ($D_0$),
the information dimension reflects the property of the dynamics on the set
through the invariant measure $p_i$.

As is explained above, the information of the saddle which characterizes
the chaotic behavior can be understood from the following quantities:
$\kappa$, $\lambda_i$, $D_0$, $D_1$ and $h_\text{KS}$.
In the following, we calculate
characteristic quantities of the fractal peak for $z_0=-1$ near
$x_0=0.45$ to confirm that what we found is transient chaos
\cite{Bleher1990}.

\begin{figure}
 \centering
 \includegraphics[width=0.5\hsize]{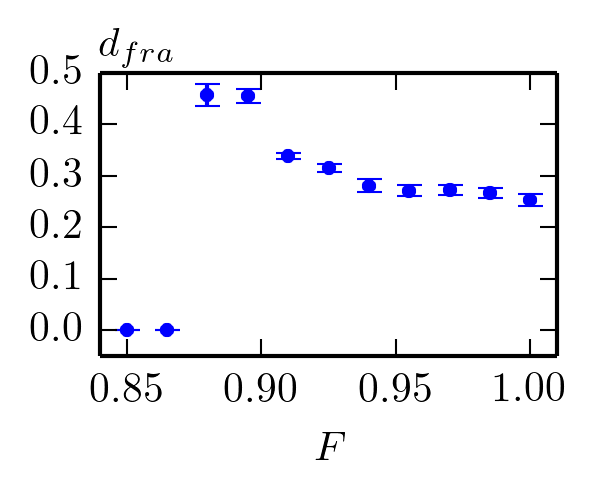}
 \caption{The abrupt bifurcation observed by varying the force term
 toward zero. $F=f\tilde{r}_c/E$ is the dimensionless force parameter,
 where $\tilde{r}_c=|q_mq_e|/\sqrt{mE}$ and $E$ is the total energy of
 the particle.
 $F=1$ corresponds to the original parameter, and
 $F=0$ corresponds to the solvable, zero force limit.
 We can see that the
 fractal dimension abruptly drops to zero around $F\sim 0.87$.}
  \label{bif}
\end{figure}
First, we calculate the uncertainty exponent as follows:
We first randomly choose
the impact parameter $x_0$ and $x_0+\epsilon$ and define that the pair
$x_0$ and $x_0+\epsilon$ is uncertain if
$|t_\text{pass}(x_0+\epsilon)-t_\text{pass}(x_0)|>0.5$
\cite{nonhyperbolic1991}. If $t_\text{pass}$ has a fractal structure,
the number of uncertain pair decays slowly as $\epsilon\to 0$ compared
to the case with the smooth structure:
If we denote the fraction of the uncertain pair among all the
randomly chosen parameter sets as $F(\epsilon)$,
$F(\epsilon)\sim \epsilon^{d_\text{unc}}$ ($\epsilon \to 0$), and
the fractal dimension
is given as
$d_\text{fra}\coloneqq 1-d_\text{unc}\coloneqq 1-\lim_{\epsilon\to 0}(\ln
F(\epsilon)/\ln\epsilon)$.
We obtained $d_\text{fra}=0.259\pm 0.008$.
We calculated this exponent for different values of 
the dimensionless force parameter $F=f\tilde{r}_c/E$,
where $\tilde{r}_c=|q_mq_e|/\sqrt{mE}$, with the total energy
$E$ fixed.
$F=1$ limit corresponds
to the original model Eq. (\ref{EoM}) with $\vec{r}(0)=(x_0,0,-\tilde{r}_c)$ and
$\vec{v}(0)=\vec{0}$,
while $F=0$ corresponds to $f=0$ in Eq. (\ref{EoM})
(i.e., the solvable limit \cite{Shnir2005,Note1})
with $\vec{r}(0)=(x_0,0,-\tilde{r}_c)$ and $\vec{v}(0)=(0,0,\sqrt{2E/m})$
\cite{Note1}.
We found that the
abrupt bifurcation similar to the potential scattering problem
\cite{Bleher1990} occurs around
$F\sim 0.87$, see Fig. \ref{bif}.

\begin{comment}
On the contrary, varying the mass parameter leads to the smooth
bifurcation. We note that by varying $m_x$, we break the rotational
symmetry, so the effective degrees of freedom becomes three, and in this
case the formula for the dimension of the crossing point mentioned in
the introduction does not hold, i.e., the saddle may have fractional
dimension but the stable manifold of it does not generically cross the
curve swept by varying $x_0$ in phase space \cite{CHEN199093}, so we cannot exclude the
possibility that the saddle still has a fractional dimension even if
$d_\text{fra}=0$.
\end{comment}

\begin{figure}
 \centering
 \includegraphics[width=\hsize]{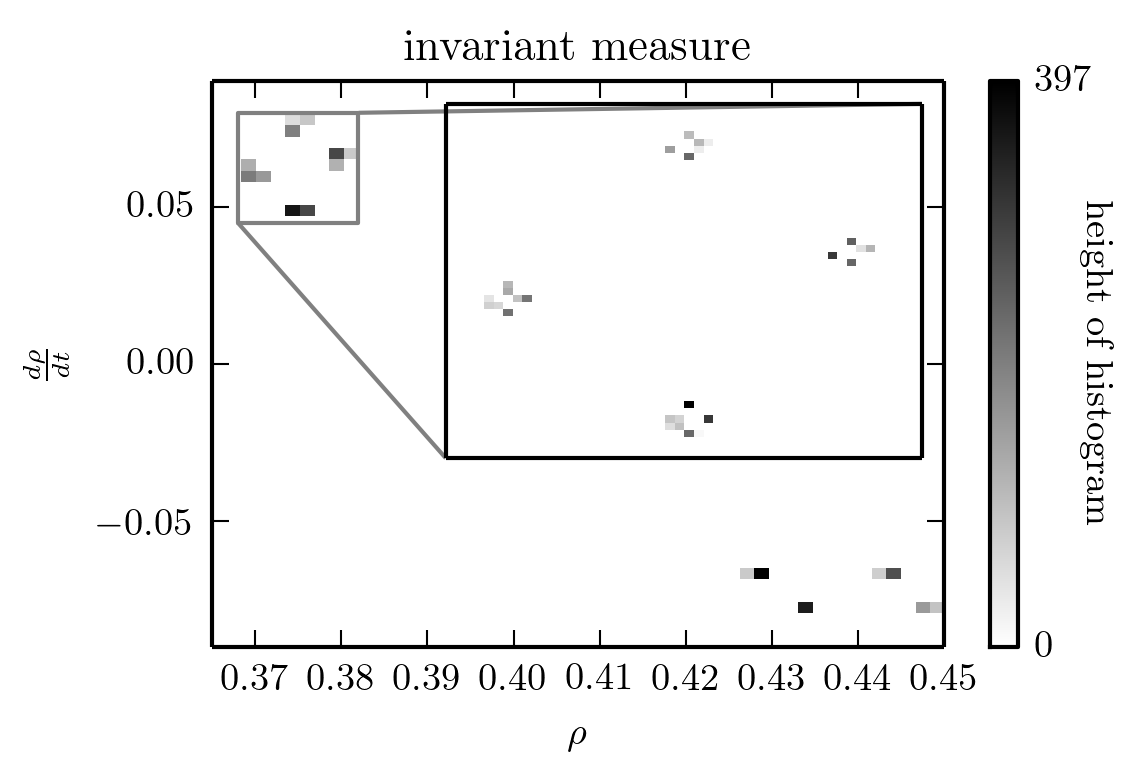}
 \caption{The invariant measure of the strange saddle on the Poincar\'e
 section defined by
 $z=-0.5 r_c$ and $v_z>0$, obtained by the PIM-triple method
 \cite{nusse1989,Bleher1990}. The number of points on Poincar\'e section
 obtained from the single trajectory is 3403, and we divide
 $(\rho,d\rho/dt)$ plane into $50\times 50$ boxes to calculate the
 histogram. The color represents the height of the histogram, as is
 shown in the right of the figure.}
  \label{pim}
\end{figure}
Secondly, we calculate the invariant measure of the strange saddle on
the Poincar\'e section defined by $z=-0.5 r_c$.
We note that our system has two degrees of freedom with the energy conservation,
Eq. (\ref{simpEoM}). Accordingly, the dimension of the Poincar\'e
section is two, and we take the coordinate as $(\rho,d\rho/dt)$, where
$\rho=\sqrt{x^2+y^2}$.
The result is shown in Fig. \ref{pim}. From this invariant
measure, we can calculate the R\'enyi dimension $D_q$ in the three
dimensional phase space, defined in
Eq. (\ref{Renyi}). We obtain
$D_2=1.51 \pm 0.02$, $D_1=1.52 \pm 0.02$ and $D_0=1.54 \pm 0.01$, so
$D_0$ and $D_1$ are almost the same within the error in our model.

Finally, we calculate the Lyapunov exponent $\lambda$ and the escape
rate $\kappa$ \cite{Note1}. We
obtained $\lambda=0.9437 \pm 0.0002$ and $\kappa=0.676 \pm 0.001$. If we
substitute these values to the Kantz-Grassberger formula with Young's formula,
we obtain the information dimension of the
saddle $D_1=1+2(1-\kappa/\lambda) = 1.568 \pm 0.002$,
which is in accordance with the result of calculated information dimension
of the saddle $D_1$.
In addition, from Young's formula, we obtain $h_\text{KS}=0.268 \pm 0.001$.
Also, the value of $D_0$ is consistent with the
uncertainty exponent by the formula
$D_0=1+2d_\text{fra}=1.52 \pm 0.02$.

Next, we discuss the stability of the chaotic behavior against the dissipative
perturbation
\cite{motter2001dissipative,seoane2006basin,seoane2007fractal}, i.e., 
$-\eta \dot{\vec{r}}$ term in the right hand side of Eq. (\ref{EoM}).
The effect of dissipation on the transient chaos has been studied and is
termed as ``doubly transient chaos''
\cite{Motter2013,Chen2017}.
The two notable features in their models
are as follows \cite{Motter2013,Chen2017}:
The dissipation leads to the exponential increase of
the escape rate, and
the fractal dimension decreases
monotonically for finer scale
($\epsilon \to 0$).
The former one is in accordance with our model \cite{Note1}, while the
latter one is in stark
contrast to our model: We found that as we look finer scale, the fractal
dimension rapidly grows and saturate at $1$ \cite{Note1}.
Since, as is discussed in
\cite{Motter2013}, the decrease of the fractal dimension as increasing
the scale is very slow, we speculate that if we look finer scale than
$10^{-10}$, the fractal dimension eventually decrease.
Also we note that this behavior may be due to the peculiarity of our
model, i.e., the presence of infinitely many backscattering
points. Indeed we observed similar kind of monotonic increase of the
fractal dimension in a slightly modified model in the absence of
dissipation \cite{Note1}, where we can attribute the increase of the
fractal dimension to the infinite number of backscattering points.
We note that this behavior strongly depends on the way we regularize the
monopole singularity, since
$r_\text{min}$ is very close to zero around the peaks \cite{Note1}.
We also found
as we further increase the dissipation, $t_\text{pass}$ becomes 
smooth, and the chaos disappears.
We further verified the stability against the perturbation which breaks
$J_z$ conservation, i.e., the deformation of the mass along $x$
direction. We calculated the fractal dimension and found that it remains
finite upon deformation \cite{Note1}. From these results, we believe
that the chaotic behavior we found can be observed in the experiments where the
various perturbations exist.

{\it Summary and Discussion.---}
We found the fractal peak structure in $t_\text{pass}$ in
the scattering
problem of a charged particle by magnetic monopole in the presence of a uniform
electric field, and verified it is the consequence of the fractal nature
of the saddle by calculating the quantities which characterize the
saddle. Although our model have a single scatter, i.e., the monopole, the
unusual nature of the scattering angle effectively leads to the
scattering problem between multiple potential hills.
We observed two bifurcation routes
(1) by varying the
electric field toward zero, where the model is solvable;
(2) by introducing the dissipation, through the intermediate region where
the fractal dimension monotonically increase as the scale becomes
finer. We clarified the stability of the chaotic behavior
against the perturbations which exist in the real experiments.
In addition to the chaotic scattering, we found that, for small
$x_0$ and $|z_0|$, the particle is captured near the monopole, i.e.,
$t_\text{pass}$ diverges.
This capturing is caused by the fact that the region which is accessible
by the particle is bounded near the monopole,
and we analytically derived the parameter
region where the capturing occurs from the form of
$U_\text{eff}$ \cite{Note1}. Since the semiclassical wave packet passing
through the diaboilic point in the Born-Oppenheimer energy bands
feels the monopole of the Berry curvature
and the dynamics of it can be modeled by Eq. (\ref{EoM}), the capturing observed
here may serve as a mechanism of the bottleneck effect where the
relaxation of the excited state toward the ground state is slowed down.

The important future work is the detailed discussion of the symmetry
breaking deformation, i.e., the mass deformation. In this case, the
system can be characterized by the fast chaotic motion and the slow mode
emerged from the small symmetry breaking perturbation.
The effect of chaotic classical system on the dynamics of
the slow degrees of freedom has been discussed and is termed as the
``geometric magnetism'' and ``deterministic friction''
\cite{berry1993chaotic,Berrymassivebil}. By extending these notions to the
transiently chaotic system with the invariant measure on the saddle
\cite{gaspard2005chaos}, it may be possible to discuss the stability
of the saddle by examining the fate of the dynamics of the slow degrees
of freedom. 

\begin{acknowledgments}
The authors thank H. Ishizuka and
X. Zhang for useful discussions.
This work was supported 
by JSPS KAKENHI Grant Number JP18J21329 (K.M.), and
JSPS KAKENHI Grant Number JP26103006, JP18H03676, and
ImPACT Program of Council for Science, Technology and
Innovation (Cabinet office, Government of Japan),
and JST CREST Grant Numbers
JPMJCR16F1, Japan (N.N.).
\end{acknowledgments}

%\bibliographystyle{apsrev4-1}
%\bibliography{reference}
 \newcommand{\noop}[1]{}
\end{document}

\title{Supplement material for ``Capture and chaotic scattering of a
charged particle by magnetic monopole
under uniform electric field''}

\author{Kou Misaki,${}^{1}$ Naoto Nagaosa${}^{1,2}$}

\affiliation{
$^1$Department of Applied Physics, The University of Tokyo, 
Bunkyo, Tokyo 113-8656, Japan
\\
$^2$RIKEN Center for Emergent Matter Science (CEMS), Wako, Saitama 351-0198, Japan
\\
}
%\date{\today}
\maketitle

\onecolumngrid
\appendix
 \section{Routhian}
 The Lagrangian describing the charged particle in the presence of the
 monopole magnetic field can be written as follows:
\begin{equation}
 L=\frac{m}{2}(\dot{\rho}^2+\dot{z}^2+\rho^2\dot{\phi}^2)-q_mq_e
  \frac{z}{\sqrt{\rho^2+z^2}} \dot{\phi}-U(\rho,z), \label{applagr}
\end{equation}
where we assumed that the potential energy $U$ preserves the axial
symmetry along $z$ direction, $(\rho,\phi,z)$ represent the cylindrical
coordinates, and the vector potential
$A_{\phi}=-q_m q_e z/\sqrt{\rho^2+z^2}$ comes from the magnetic monopole
at the origin. Since $\phi$ is the cyclic coordinate of the system, the
conjugate momentum
\begin{equation}
 J_z=\frac{\partial L}{\partial \dot{\phi}}=m\rho^2\dot{\phi}-q_m q_e
  \frac{z}{\sqrt{\rho^2+z^2}}
\end{equation}
is conserved. Then, we can construct the Routhian \cite{goldsteinmech}:
\begin{equation}
 R=L-J_z \dot{\phi}|_{\dot{\phi}=(J_z+q_m q_e
  \frac{z}{\sqrt{\rho^2+z^2}})/(m\rho^2)}
  =\frac{m}{2}(\dot{\rho}^2+\dot{z}^2)-\frac{1}{2m\rho^2}
  \left(J_z+q_m q_e \frac{z}{\sqrt{z^2+\rho^2}}\right)^2-U(\rho,z).
\end{equation}
As we can see, the dynamics is now described by two degrees of
freedom $(\rho,z)$, and the potential energy is modified to
be
\begin{equation}
 U_\text{eff}(\rho,z)\coloneqq U(\rho,z)+\frac{1}{2m\rho^2}
  \left(J_z+q_m q_e \frac{z}{\sqrt{z^2+\rho^2}}\right)^2. \label{Ueffapp}
\end{equation}

\section{The trapping region of $U_\text{eff}$}
\begin{figure}
 \centering
 \includegraphics{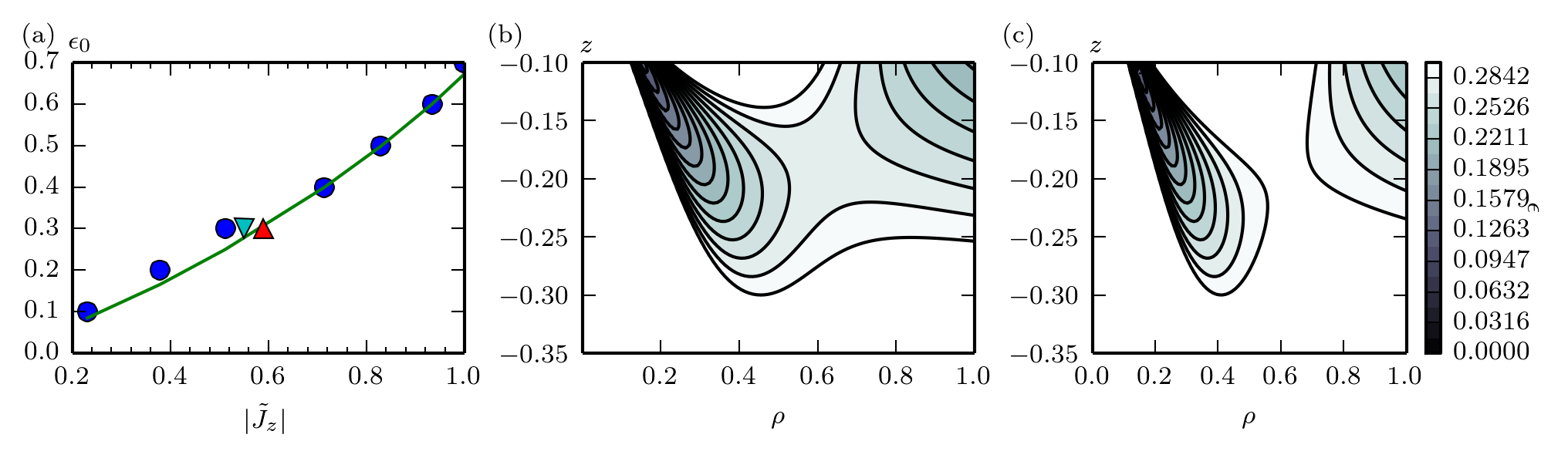}
 \caption{(a) The boundary separating the trapping orbit and the escaping
 orbit. The dots are the result of the numerical calculation and the
 line represents the analytic expression Eq. (\ref{saddleboundary}).
 (b,c) The contour plot for the energy range $0<\epsilon<\epsilon_0=0.3$
 and $\tilde{J}_z=0.55$ and $0.59$ for (b) and (c), respectively.
 The parameter values for (b) and (c) are shown
 by the blue down triangle and red up triangle
 in $(a)$, respectively.}
  \label{saddleposronbun}
\end{figure}

As is discussed in the main text, if we vary the two parameters
$(x_0,z_0)$, $J_z(x_0,z_0)$ and $E_0(z_0))$ change and for small $E_0$ and
large $J_z$, the region with $E\leq E_0$ is separated and confined to
the finite region, as we can see from Fig. \ref{saddleposronbun}(c),
thereby leading to the infinite $t_\text{pass}$.
For smaller $J_z$, the
two regions are connected by the saddle with the energy height $E_\text{saddle}<E_0$,
leading to finite $t_\text{pass}$. To obtain the analytical expression
of the boundary between these two cases in $(J_z,E_0)$ plane,
we observe that,
\begin{align}
 U_\text{pass}=E_0 &\Leftrightarrow -Z+\frac{1}{2P^2}
  \left(\frac{Z}{\sqrt{Z^2+P^2}}
 +\tilde{J}_z\right)^2=\epsilon_0 \nonumber\\
 &\Leftrightarrow 2\sin v\cos^2 v R^3 +2\epsilon_0 \cos^2 v
 R^2-(\sin v+\tilde{J}_z)^2=0,
\end{align}
where $Z\coloneqq r_c z$, $P\coloneqq r_c \rho$, $R\coloneqq r_c r$,
$Z=R\sin v$, $P=R\cos v$, $r_c=|q_mq_e|^{2/3}/(mf)^{1/3}$,
$\tilde{J}_z\coloneqq J_z/(q_mq_e)$ and
$\epsilon_0\coloneqq E_0/(f r_c)$. The necessary condition for the
merger of two separate region is that the discriminant for this third
order polynomial for $R$ is zero:
\begin{equation}
 \sin^4 v+2\tilde{J}_z\sin^3
  v+\left(\tilde{J}_z^2+\alpha\right)\sin^2
  v-\alpha=0,
\end{equation}
where $\alpha\coloneqq 8\epsilon_0^3/27$.
Furthermore, we impose the condition that the discriminant for this
fourth order polynomial for $\sin v$ is zero:
\begin{equation}
 \alpha^2(\alpha^3+(3\tilde{J}_z^2+8)\alpha^2+(3\tilde{J}_z^4-20\tilde{J}_z^2+16)\alpha
 +\tilde{J}_z^4(\tilde{J}_z^2-1))=0.
\end{equation}
Solving this equation for $\alpha$, we obtain five solutions. Choosing
the relevant solution, we obtain
\begin{equation}
 \epsilon_0=\left(-\frac{9}{8}(8+3\tilde{J}_z)+
			 \frac{9}{2}\frac{4+27\tilde{J}_z^2}{G}
			      +\frac{9}{8} G\right)^{\frac{1}{3}},
 \label{saddleboundary}
\end{equation}
where
\begin{equation}
 G\coloneqq \left(64-1080\tilde{J}_z^2-\frac{729
						      \tilde{J}_z^4}{2}+\frac{3\sqrt{3}}{2}\sqrt{\tilde{J}_z^2(-32+27\tilde{J}_z^2)^3}\right)^{\frac{1}{3}}.
\end{equation}
The plot of this curve is shown by the solid line in
Fig. \ref{saddleposronbun}(a). As a bonus, we obtain the value of $J_z$
where the saddle point of the potential energy vanishes:
Above $\tilde{J}_z^2=32/27$, the right hand side of
Eq. \ref{saddleboundary} becomes complex, meaning that there is no saddle
in the potential energy. Since the chaotic scattering is caused by the
pocket of energy minimum bounded by the saddle, we expect no chaotic
scattering above this $J_z$.

\section{Dynamics of charged particle in the presence of monopole}
\begin{figure}
 \centering
 \includegraphics[width=\hsize]{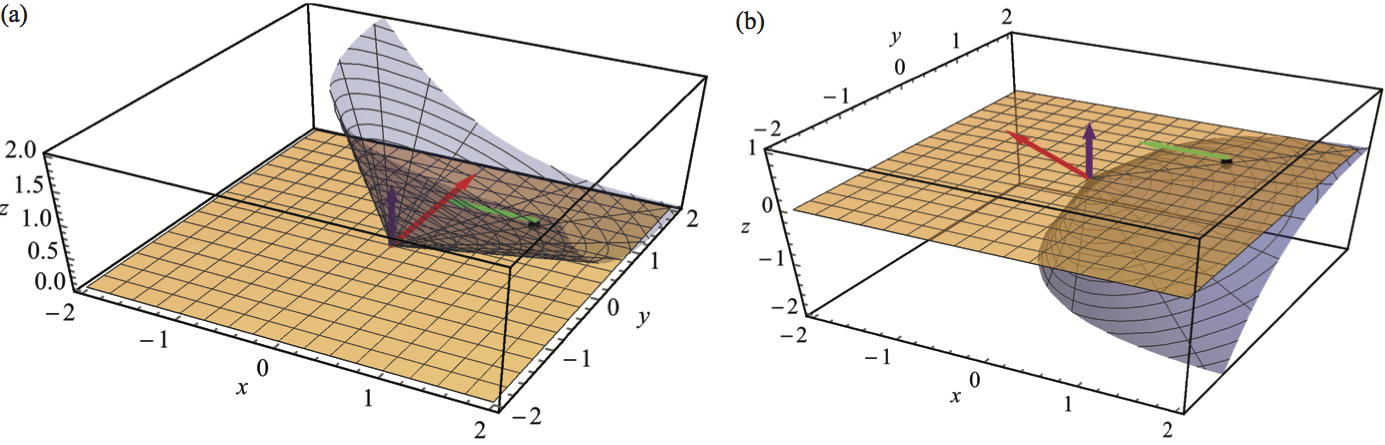}
 \includegraphics[width=0.2\hsize]{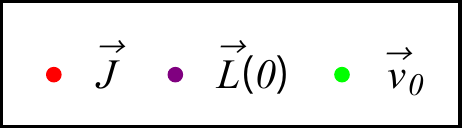}
 \caption{The cone on which the trajectory of the particle
 lies (shown blue), in the case of (a) $q_eq_m<0$ and (b) $q_eq_m>0$. We can see
 the direction of the tilting of $\vec{J}$ and the direction of the
 bending of $x-y$ plane into the cone differ in the two cases.
 The black dot represents the initial position of the
 particle, and the red, purple and green arrows represent $\vec{J}$,
 $\vec{L}(0)$ and $\vec{v}_0$, respectively. We took the $z$ axis along
 $\vec{L}(0)$ direction, which coincides with $\vec{J}$ in the case of
 $q_eq_m=0$, and the orange plane represents $x-y$ plane.} 
  \label{cone}
\end{figure}

 Here, we review the analytic solution in the case of $f=0$
 \cite{Darboux1878,poincare1896remarques}, for completeness.
 The equation of motion is,
\begin{equation}
 m\ddot{\vec{r}}=q_eq_m
 \frac{\dot{\vec{r}}\times\vec{r}}{r^3}. \label{monceq}
\end{equation}
By taking the inner product with $\dot{\vec{r}}$, and the outer product
with $\vec{r}$, we obtain four
conserved quantities:
\begin{equation}
 E\coloneqq\frac{m}{2}(\dot{\vec{r}})^2,\quad
  \vec{J}\coloneqq m\vec{r}\times \dot{\vec{r}}-q_eq_m
  \frac{\vec{r}}{r}\eqqcolon \vec{L}-q_eq_m \frac{\vec{r}}{r}. \label{conscone}
\end{equation}
Among them, the independent conserved quantities are $E$, $J_z$ and $|\vec{J}|$.
If we take the inner product with $\vec{r}$ and Eq. (\ref{monceq}), we obtain
$\vec{r}\cdot\ddot{\vec{r}}=0$, so
\begin{equation}
 \frac{d^2 r^2}{dt^2}=2 (\dot{\vec{r}})^2=\frac{4E}{m}, \quad \therefore \,
  r(t)=\sqrt{(\vec{v}_0)^2t^2+2 (\vec{r}_0\cdot
  \vec{v}_0)t+(\vec{r}_0)^2}=|\vec{v}_0t+\vec{r}_0|, \label{solr}
\end{equation}
where $\vec{r}(0)=\vec{r}_0$ and $\dot{\vec{r}}(0)=\vec{v}_0$.
We assume that $\vec{r}_0$ is not parallel to $\vec{v}_0$.
Here, we
take $\vec{J}$ along the $z$ direction and take the spherical
coordinate $(r,\theta,\phi)$. Then,
\begin{equation}
 \frac{\vec{J}\cdot\vec{r}}{r}=-q_eq_m
  \Leftrightarrow \cos\theta=\frac{-q_eq_m}{J}
  =\frac{-q_eq_m}{\sqrt{L^2+(q_eq_m)^2}}. \label{solth}
\end{equation}
 Finally, we take the inner
product of $\vec{J}$ and $\vec{L}$:
\begin{equation}
 \vec{J}\cdot \vec{L}=J^2-(q_eq_m)^2
  \Leftrightarrow L_z= J\sin^2\theta.
\end{equation}
Since $L_z=m r^2\sin^2\theta \dot{\phi}$, we obtain
\begin{equation}
 \dot{\phi}=\frac{J}{mr^2}, \quad \therefore \,
  \phi(t)=\frac{1}{\sin\theta}\left[\tan^{-1}\left(\frac{(\vec{v}_0)^2 t+
				     \vec{r}_0\cdot
				     \vec{v}_0}{|\vec{r}_0\times
				     \vec{v}_0|}\right)-\tan^{-1}\left(
				     \frac{\vec{r}_0\cdot\vec{v}_0}{|\vec{r}_0\times
				     \vec{v}_0|}
								 \right)
		     \right] \label{solphi}
\end{equation}
Combining Eqs. (\ref{solr}), (\ref{solth}) and (\ref{solphi}), we
obtained the analytical solution.
From Eq. (\ref{solth}), we can see that the the motion of the particle
is restricted on the cone
oriented along $\vec{J}$ with the angle $\theta_0=\cos^{-1}(-q_eq_m/J)$.
Here we note that the motion is similar to the motion in the absence of
the monopole in two aspects: the motion is restricted to the two
dimensional plane (i.e., the cone), and the trajectory on that plane is
geodesic. To see the first
aspect in more detail, we fix $\vec{r}_0$ and $\vec{v}_0$ and show the
form of the cone in Fig.\ref{cone}.
Here, we let $\vec{L}(0)$ along the $z$ axis to show the difference between
$q_eq_m<0$ and $q_eq_m>0$ cases (Note the difference of the coordinate system
from the analytic solution.).
We consider the process where we start from the
$q_eq_m=0$ and gradually increase $|q_eq_m|$. That results in the two
modifications: The direction of the total angular momentum tilts toward
the position of the particle, $\vec{r}_0$, and the surface where the
trajectory of the particle lies is bent from the $x-y$ plane into the
cone. We note that, since the $x-y$ plane is tangential to the cone at
$\vec{r}_0$ and the initial velocity is along the $x-y$ plane, the
particle will not get out of the cone \footnote{In other words, once the
direction of $\vec{J}$ is fixed from Eq. (\ref{conscone}), then the
opening angle is automatically determined from the requirement that the
cone should be tangential to the plane subtended by the vectors
$\vec{r}_0$ and $\vec{v}_0$.}.
Since the trajectory of the particle at $q_eq_m=0$ is the geodesic
on the $x-y$ plane, we expect the motion of the particle on the cone is
also the geodesic.
To verify that this is correct, we define $\chi$ as the angle between
$\vec{r}_0$ and $\vec{v}_0$. Then,
\begin{equation}
 \frac{\vec{r}_0\cdot \vec{v}_0}{|\vec{r}_0\times \vec{v}_0|}=
 \cot \chi=
\tan
  \left(\frac{\pi}{2}-\chi\right),
\end{equation}
so from Eq. (\ref{solphi}),
\begin{align}
 &\frac{(\vec{v}_0)^2 t+\vec{r}_0\cdot \vec{v}_0}{|\vec{r}_0\times
  \vec{v}_0|}
  =\tan\left(\sin\theta \phi(t)+\frac{\pi}{2}-\chi\right)
 =-\cot (\sin\theta\phi(t)-\chi) \nonumber\\
 \therefore \, & \frac{(\vec{v}_0)^2 r(t)^2+(\vec{r}_0\cdot \vec{v}_0)^2-(\vec{v}_0)^2(\vec{r}_0)^2}{|\vec{r}_0\times
  \vec{v}_0|^2}=\cot^2 (\sin\theta\phi(t)-\chi)\quad (\because
 \text{Eq. (\ref{solr})})\nonumber\\
 \Leftrightarrow &\frac{(\vec{v}_0)^2 r(t)^2}{|\vec{r}_0\times
 \vec{v}_0|^2}=\cot^2 (\sin\theta\phi(t)-\chi)+1 \nonumber\\
  \therefore \, & r(t) \sin(\chi-\sin\theta\phi(t))=r_0 \sin\chi.
\end{align}
To understand this equation, we expand the cone to obtain the
development, and consider the two dimensional motion on that.
Then, the quantity $\sin\theta\phi(t)$ represents the angle between
$\vec{r}_0$ and $\vec{r}(t)$ in the development of the cone. Since $\chi$
represents the angle between $\vec{r}_0$ and $\vec{v}_0$, the above
equation can be rewritten as:
\begin{equation}
 \vec{v}_0\cdot\vec{r}(t)=\vec{v}_0\cdot\vec{r}_0 \quad \text{on the
  development of the cone.}
\end{equation}
It means that the motion of the particle is a straight line on the
development. In other words, the trajectory is geodesic for the flat metric
induced by (locally) identifying the development of the cone with the two dimensional
Euclidean space. In this sense, the trajectory is still ``straight'',
although in three dimensional space it looks complicated.

One important feature which is relevant for the main text is
Eq. (\ref{solr}). It means that the monopole does not attract the
particle to cause the delay of the passing of the particle near the
origin.

\section{Hamilton-Jacobi equation}
Here we solve $f=0$ case with the Hamilton-Jacobi method, which may serve
as a good starting point for further analysis with the perturbation theory.
The Hamiltonian for Eq. (\ref{applagr}) obtained by the Legendre
transformation is the following:
\begin{equation}
 H=\frac{p_r^2}{2m}+\frac{1}{2mr^2}\left[p_{\theta}^2+\frac{1}{\sin^2\theta}\left(p_{\phi}+q_mq_e
									    \cos\theta\right)^2\right].
\end{equation}
The Hamilton-Jacobi equation for this Hamiltonian is
\begin{equation}
 \frac{\partial S}{\partial t}+H\left(q_i,\frac{\partial S}{\partial
				 q_i}\right)=0\Leftrightarrow \frac{\partial S}{\partial t}+\frac{1}{2m}\left(\frac{\partial S}{\partial
  r}\right)^2+\frac{1}{2mr^2}\left[\left(\frac{\partial S}{\partial
		     \theta}\right)^2+\frac{1}{\sin^2\theta}\left(\frac{\partial S}{\partial \phi}+q_mq_e
									    \cos\theta\right)^2\right]=0.
\end{equation}
Since the energy is conserved, $\phi$ is the cyclic coordinate, and $r$
and $\theta$ are separable, we put $S=-\epsilon
t+W_r(r)+W_{\theta}(\theta)+\alpha_{\phi}\phi$ to obtain
\begin{align}
 &\left(\frac{\partial W_\theta}{\partial
		     \theta}\right)^2+\frac{1}{\sin^2\theta}\left(\alpha_\phi+q_mq_e
 \cos\theta\right)^2 =\alpha_\theta^2-(q_eq_m)^2, \\
 &\left(\frac{\partial W_r}{\partial
  r}\right)^2+\frac{\alpha_\theta^2-(q_eq_m)^2}{r^2}=2m\epsilon,
\end{align}
where $\alpha_\theta$ and $\alpha_\phi$ corresponds to $|\vec{J}|$ and
$J_z$, respectively. At this point, $\epsilon,\alpha_\theta$, and
$\alpha_\phi$ are constants of motion.
These equations can be transformed to
\begin{align}
 W_\theta&=-\int^\theta d\theta
 \sqrt{\alpha_\theta^2-(q_eq_m)^2-\frac{1}{\sin^2\theta}(q_eq_m\cos\theta+\alpha_\phi)^2},\\
 W_r &= \int^r dr \sqrt{2m\epsilon -\frac{\alpha_\theta^2-(q_eq_m)^2}{r^2}},
\end{align}
where we chose the minus sign for $W_\theta$ just for convenience.
Then, we calculate $\beta_i(q_i,\alpha_i) \coloneqq \frac{\partial S}{\partial
\alpha_i}$, which are canonically conjugate to $\alpha_i$ and are the remaining constants
of motion. After a bit long but straightforward calculation, we obtain
\begin{align}
 \beta_\phi &\coloneqq \frac{\partial S}{\partial \alpha_\phi}=\phi - \frac{\pi}{2}+
 \arctan\left[\frac{\cos A\sin i-\cos i \sin A\sin
 \psi}{\cos \psi \sin A}\right], \label{HamJac1}\\
 \beta_\theta&\coloneqq \frac{\partial S}{\partial \alpha_\theta}=
 \frac{1}{\sin A}\left(\frac{\pi}{2}-\arctan
 \left[\sqrt{\frac{2m\epsilon r^2}{\alpha_\theta^2\sin^2A}-1}\right]\right)
 + \psi,\label{HamJac2}\\
 \beta_0&\coloneqq \frac{\partial S}{\partial \epsilon}=-t +
 \sqrt{\frac{m}{2\epsilon}}\sqrt{r^2-\frac{\alpha_\theta^2\sin^2 A}{2m\epsilon}}, \label{HamJac3}
\end{align}
where
\begin{align}
 \cos\theta&=\sin A \sin i \sin \psi+\cos A \cos i,\\
 \cos A&\coloneqq\frac{-q_eq_m}{\alpha_\theta},\\
 \cos i &\coloneqq \frac{\alpha_\phi}{\alpha_\theta}.
\end{align}
By solving Eqs. (\ref{HamJac1}), (\ref{HamJac2}) and (\ref{HamJac3}) for
$(r,\psi,\phi)$, noting that we are free to add $\Delta S(\alpha_i)$ to
$S$ to shift each $\beta_i$ by any function of $\alpha_i$, we obtain
\begin{align}
 r(t)&=\sqrt{\frac{2\epsilon}{m}(t+\beta_0)^2+\frac{\alpha_\theta^2\sin^2
 A}{2m
 \epsilon}}, \label{soljacr}\\
 \psi(t)&=\beta_\theta+\frac{1}{\sin
 A}\arctan\left[\frac{2\epsilon}{\alpha_\theta \sin
 A}(t+\beta_0)\right], \label{soljacpsi}\\
 \phi(t)&=\beta_\phi-\arctan\left[\frac{\cos A\sin i-\cos i \sin A\sin
 \psi(t)}{\cos \psi(t) \sin A}\right]. \label{soljacphi}
\end{align}
Note that if we substitute
\begin{align}
 \beta_0&=\frac{\vec{v}_0\cdot\vec{r}_0}{(v_0)^2},\\
 \beta_\theta &=-\frac{1}{\sin
 A}\arctan\left[\frac{2\epsilon \beta_0}{\alpha_\theta \sin
 A}\right],
\end{align}
into Eqs. (\ref{soljacr}) and (\ref{soljacpsi}), we reproduce Eqs. (\ref{solr}) and (\ref{solphi}).

\section{On the peak structure of $t_\text{pass}$}
To understand the peak structure of $t_\text{pass}$, we consider the
following toy model which can be smoothly deformed to our model: We put
the potential barrier at $z=-A<0$ plane, and consider the motion of the
particle bouncing back and force between this potential barrier and the
magnetic monopole at the origin. We start from the limit where the
barrier is step-like and far from the origin, and smoothly vary the
position and thickness of the barrier to reach the limit where the
potential can be regarded as the one imitating the uniform electric
field.
\begin{figure}
 \centering
 \includegraphics{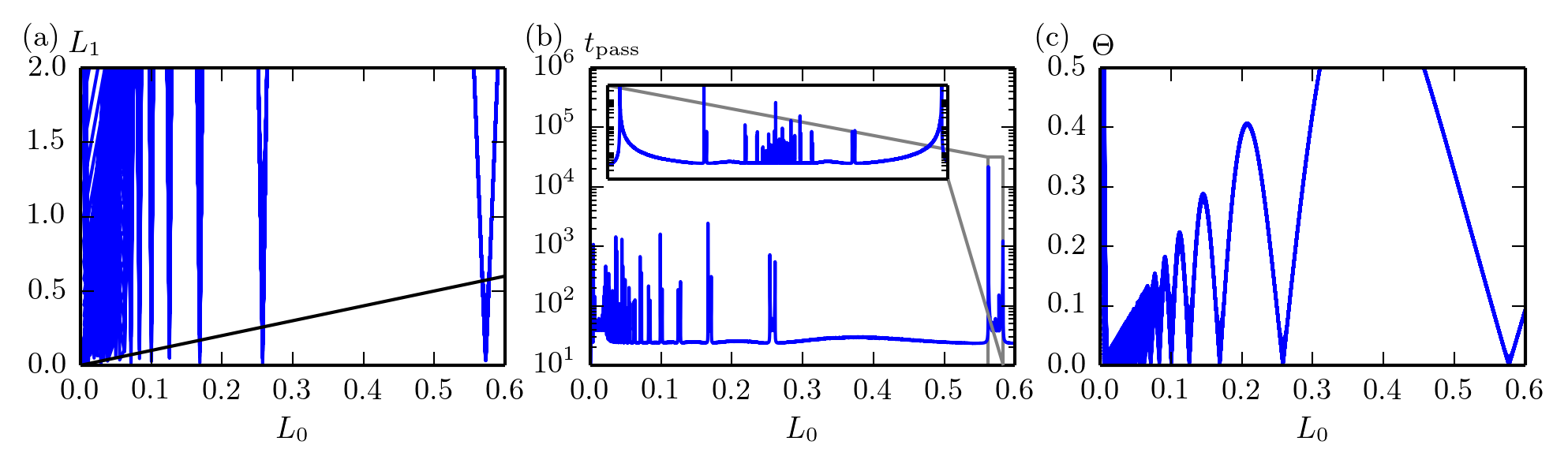}
 \caption{$L_1$, $t_\text{pass}$ and $\Theta$ as a function of $L_0$ at
 $A=B=10$.
 As we can see, the peak of $t_\text{pass}$ is near the region where
 $L_0=L_1$ holds.}
  \label{apppos10haba10}
\end{figure}
\begin{figure}
 \centering
 \includegraphics{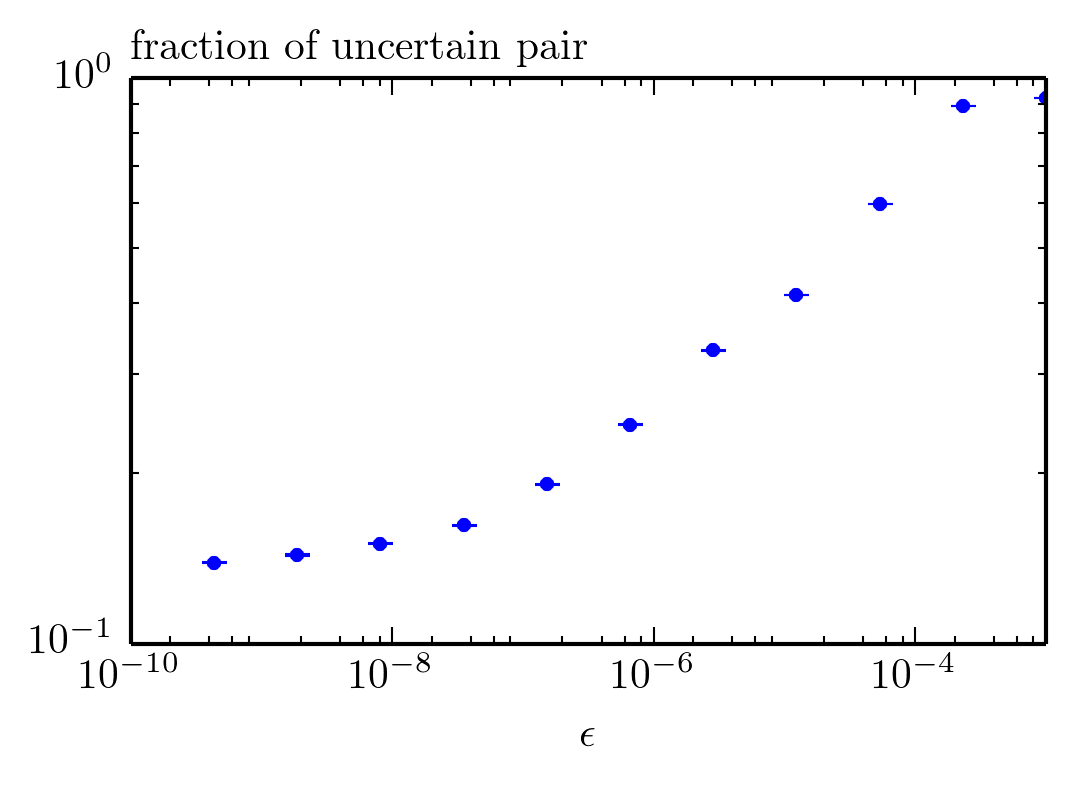}
 \caption{The fraction of uncertain pair.
As we can see, the uncertainty exponent $d_\text{unc}$ given by the
 slope (the fractal dimension given by $d_\text{fra}=1-d_\text{unc}$) decreases
 (increases) as we see finer scale ($\epsilon\to 0$).
 The calculation was done for the peak around $L_0=0.6$ with $A=B=10$, see
 Fig. \ref{apppos10haba10}(b).}
  \label{tanhtestunc}
\end{figure}

More concretely, we consider the potential of the form
$U(z)=1-\tanh((z+A)/B)$, where $A$ represents the position of the wall
and $B$ represents the width of the slope of the potential. We set the
initial condition as $\vec{v}_0=\vec{0}$ and $(x_0,y_0,z_0)=(0,0,-A)$,
and calculated $L=|\vec{r}\times \vec{v}|$ and $v_z$ at the first two
intersections with the Poincar\'e section at $z=-A+1$ with $v_z>0$. We
note that, because of $J_z$ and $E$ conservation, the effective phase space
degrees of freedom is three, so the Poincar\'e map is defined as
$(L_n,v_{z,n})\to (L_{n+1},v_{z,n+1})$. We show $L_1$, $t_\text{pass}$
and $\Theta \coloneqq \arccos(\vec{v}_{0}\cdot\vec{v}_{1})$ as a
function of $L_0$ in Fig. \ref{apppos10haba10}. As we can see,
complicated peak structure is observed near the point where $L_0=L_1$
holds. We note that near these regions, $\Theta \sim 0$, so we speculate
that the Poincar\'e map can be approximated as $(L_n)\to (L_{n+1})$ at
least for this initial regime. By considering the web diagram in
Fig. \ref{apppos10haba10}(a), we can understand the complicated
peak structure of Fig. \ref{apppos10haba10}(b):
The particles which start from the
points sandwiched between the crossing points of blue and black lines
goes to smaller $L$ region, and among them, some particles again get to the
region sandwiched between the crossing points and mapped to smaller $L$
region, and so on. Therefore, we conclude that the complicated peak
structure of $t_\text{pass}$ (Fig. \ref{apppos10haba10}(b))
and the monotonic increase of the fractal
dimension as a function of scale (Fig. \ref{tanhtestunc})
is caused by infinite number of
backscattering points, around which $L$ is mapped to smaller value.
\begin{figure}
 \centering
 \includegraphics{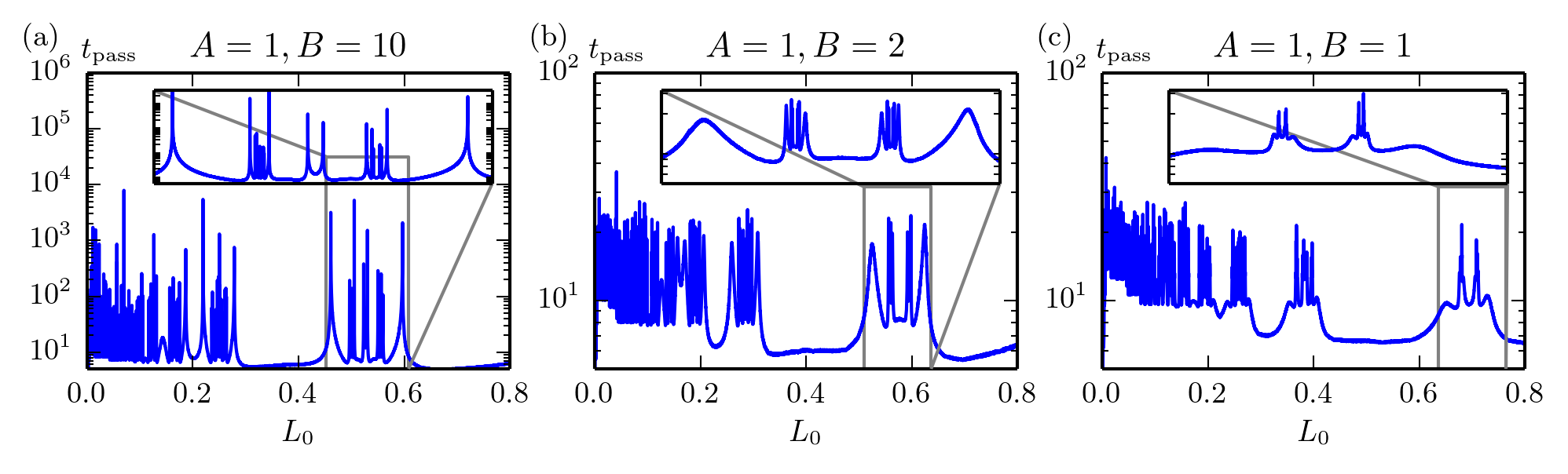}
 \caption{$L_1$, $t_\text{pass}$ and $\Theta$ as a function of $L_0$.
 As we can see, the peak of $t_\text{pass}$ is near the region where
 $L_0=L_1$ holds.}
  \label{tpasspeakapp}
\end{figure}

As we deform $A$ and $B$ to reach the model which is similar to our
model with uniform electric field, these peak structure evolves into the
fractal peak, see Fig. \ref{tpasspeakapp}. So, from the argument above,
we speculate that the multiple backscattering points plays the role of
the potential hills in the usual chaotic scattering
\cite{Gaspard1989,Bleher1990}, thereby producing the infinite number of
symbol sequence by labeling each backscattering points by integers.

\section{The change of $U_\text{eff}$ from the variation of $f$}
\begin{figure}
 \centering
 \includegraphics[width=\hsize]{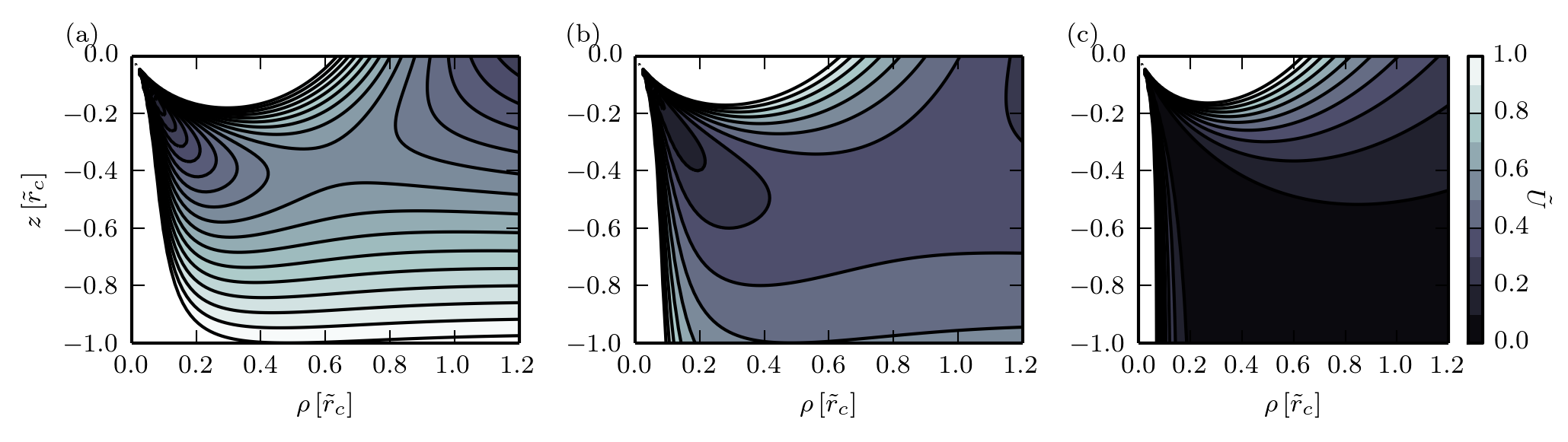}
 \caption{The change of $U_\text{eff}$ by varying the dimensionless
 force parameter, $F$. Here we set
 $J_z/(q_mq_e)=0.9$, and (a) $F=1$, (b) $F=0.5$, (c) $F=0$.}
  \label{forcesadronbun}
\end{figure}
We consider the one parameter family which connects
the equation of motion for Eq. (\ref{Ueffapp}) where $U(\rho,z)=-fz$ to the one 
where $U(\rho,z)=0$, the solvable limit.
We define $\tilde{r}_c=|q_mq_e|/\sqrt{mE}$ and
$\tilde{t}_c=|q_mq_e|/E$. Then the equation of motion can be rewritten as
\begin{align}
  &\left\{
 \begin{alignedat}{2}
  \frac{d^2 \tilde{Z}}{d\tilde{\tau}^2}&=-\frac{\partial \tilde{U}_\text{eff}}{\partial \tilde{Z}}\\
  \frac{d^2 \tilde{P}}{d\tilde{\tau}^2}&=-\frac{\partial \tilde{U}_\text{eff}}{\partial \tilde{P}},
  \end{alignedat}
 \right.
 \quad
 \tilde{U}_\text{eff}\coloneqq -F \tilde{Z}+
 \frac{1}{2}\left(\frac{J_z}{q_mq_e}+ \frac{\tilde{Z}}{\sqrt{\tilde{Z}^2+\tilde{P}^2}}\right)^2, 
 \label{simpEoMapp}
\end{align}
where $\tilde{Z}=z/\tilde{r}_c$, $\tilde{P}=\rho/\tilde{r}_c$,
$\tilde{\tau}=t/\tilde{t}_c$ and $F=f\tilde{r}_c/E$. The change of
$U_\text{eff}$ as a function of $F$ is shown in Fig. \ref{forcesadronbun}.

\section{The calculation of the Lyapunov exponent and the escape rate}
\begin{figure}
 \centering
 \includegraphics[width=\hsize]{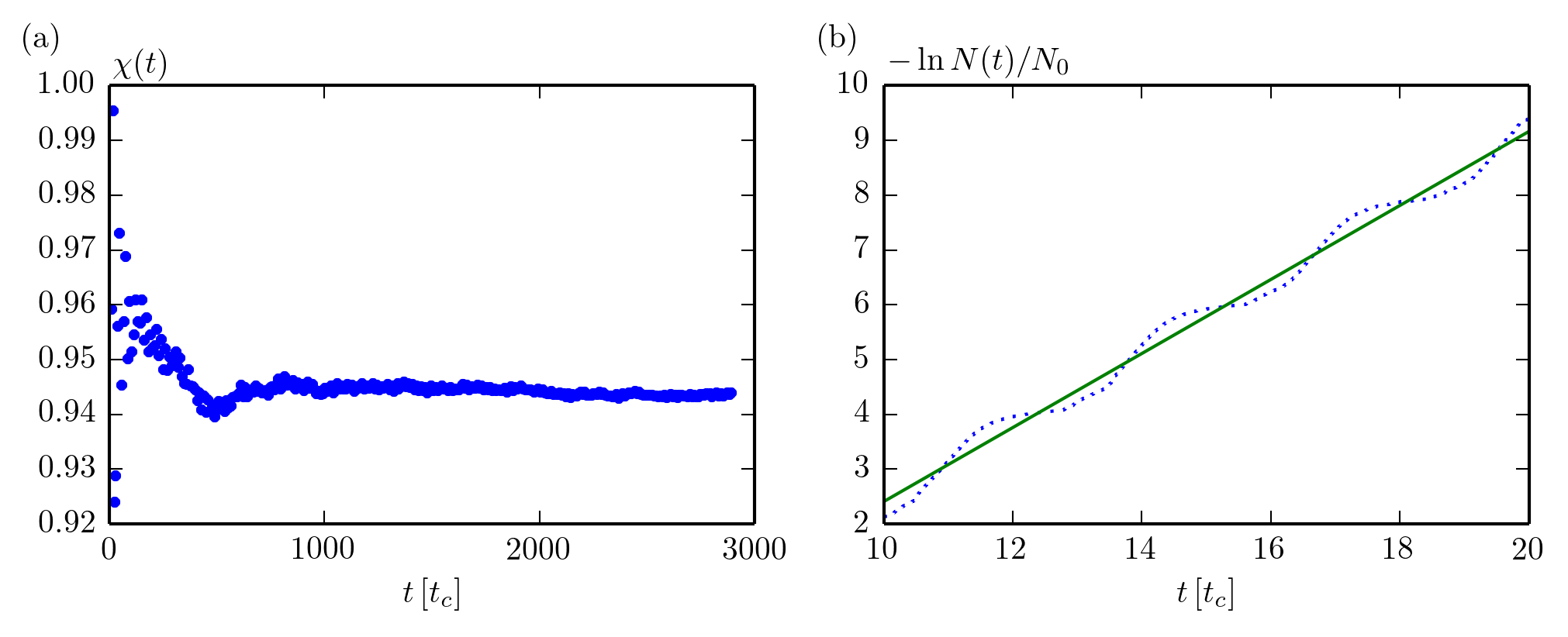}
 \caption{The calculation of the (a) Lyapunov exponent and (b) escape
 rate. (a) We calculate $\chi(t)$ by evolving the two nearby
 phase space points and measure the deviation $\alpha_i$ after the time
 $\tau_i$ to obtain
 $\chi(t)=(1/\sum_i \tau_i)\sum_i \alpha_i$, of which we obtain the Lyapunov
 exponent from the limiting value \cite{Bleher1990,ott2002}. (b) The
 dotted line represents the numerically obtained number of surviving
 particles as a function of time, and the solid line is the fitting
 curve. The slope of the curve is the escape rate.}
  \label{lyapunovescape}
\end{figure}
We calculated the Lyapunov exponent from the trajectory obtained by the
PIM-triple method, and the escape rate from the time evolution of the
number of the surviving particles. The result is shown in
Fig. \ref{lyapunovescape}. The limiting value of $\chi(t)$ in
Fig. \ref{lyapunovescape}(a) gives us the Lyapunov exponent, and the slope
of the fitting curve (the solid line) in Fig. \ref{lyapunovescape}(b)
represents the escape rate $\kappa$. We note that, since our system is
the Hamiltonian system with two degrees of freedom, the energy
conservation leads to zero Lyapunov exponent, and the direction of the
flow corresponds to another direction with zero Lyapunov exponent. In
addition, since the symplectic structure leads to the symmetric
distribution of the Lyapunov exponents around zero, the calculation of
the one positive Lyapunov exponent is enough.

\section{$t_\text{pass}$, $r_\text{min}$ and the fractal dimension in
 the presence of the dissipation}
\begin{figure}
 \centering
 \includegraphics[width=\hsize]{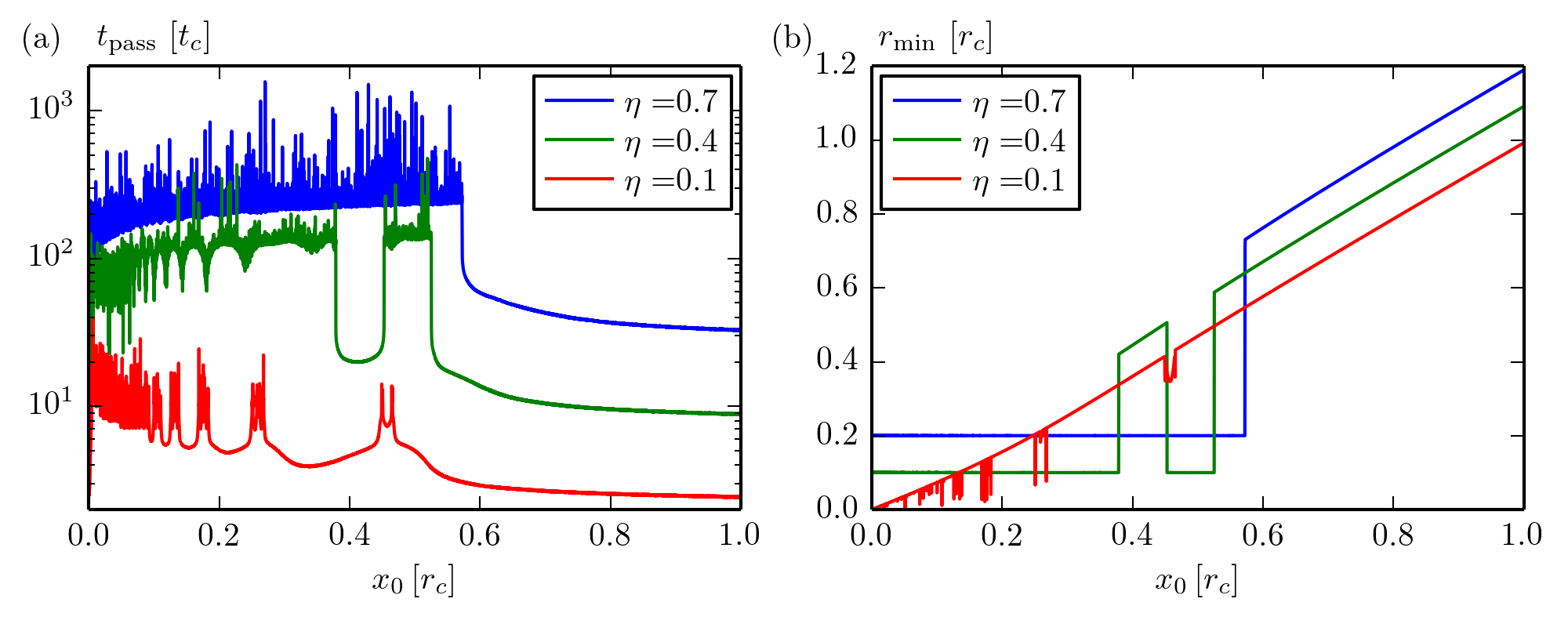}
 \caption{The plots of the (a) $t_\text{pass}$ and (b) $r_\text{min}$.
 The value of $\eta$ is shown above (a). As we can see, as we
 increase the strength of the dissipation, the fractal peak becomes
 broad with rapid variation and $r_\text{min}$ is very close to zero,
 which means that this behavior strongly depends on the way we regularize
 the monopole singularity. Note that the plots are shifted upward.}
  \label{dissdoubtpass}
\end{figure}
\begin{figure}
 \centering
 \includegraphics[width=\hsize]{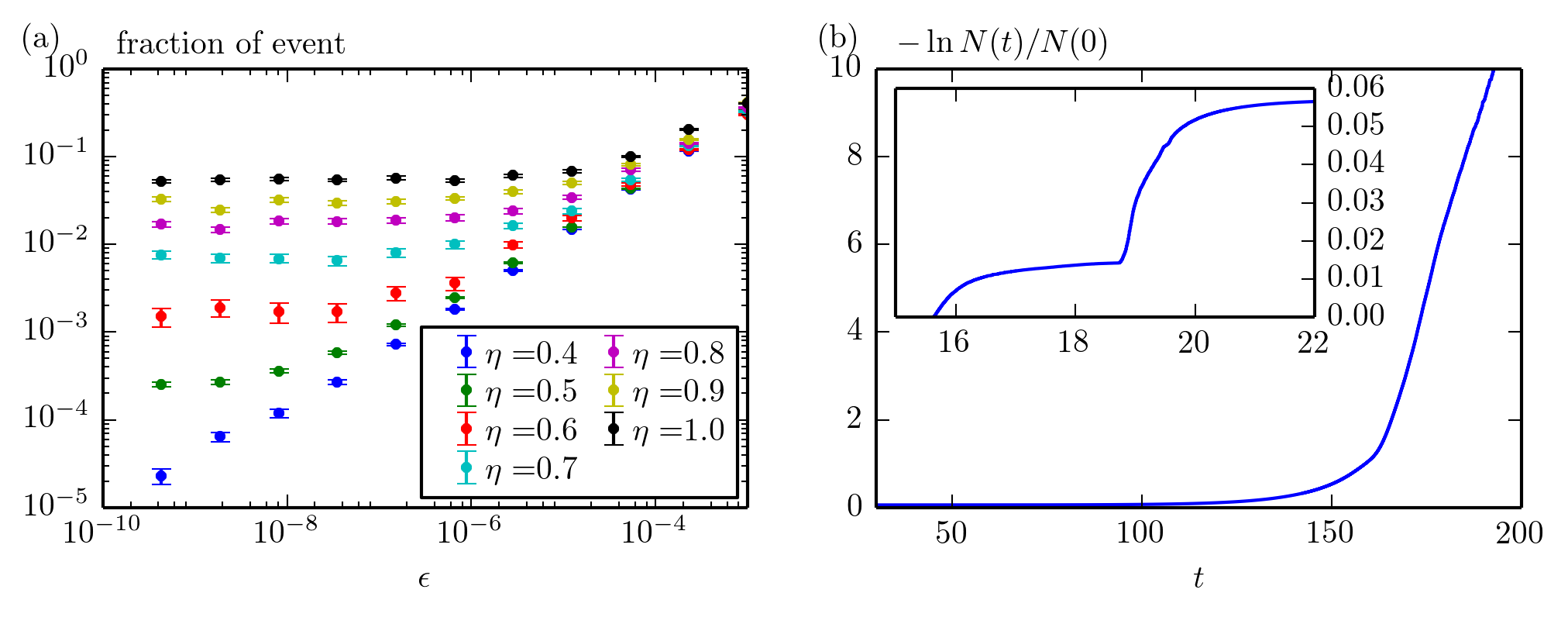}
 \caption{The plots of the (a) fraction of uncertain pair,
 and (b) the minus logarithm of the number of surviving
 particles.
 (a) As we can see, the uncertainty exponent $d_\text{unc}$ given by the
 slope in
 $(a)$ (the fractal dimension given by $d_\text{fra}=1-d_\text{unc}$) decreases
 (increases) as we see finer
 scale ($\epsilon\to 0$),
 which is in stark contrast to the doubly transient chaos.
 The value of $\eta$ is as is shown in the legend.
 (b) The inset shows the early time behavior, which shows linear behavior in
 time. After some time (around $t=100$),
 the number of escaping particles suddenly increases, which means the
 monotonic increase of the escape rate.}
  \label{dissdoubtrans}
\end{figure}
We introduced the friction term $-\eta \dot{\vec{r}}$ in our model to quantify
the stability of our model against dissipation. We calculated
$t_\text{pass}$ and $r_\text{min}$, see Fig. \ref{dissdoubtpass}. As we
can see, upon increasing the amount of the dissipation, the peak
structure of $t_\text{pass}$ drastically changes, and the fractal
dimension shows monotonic increase as we see finer scale
($\epsilon\to 0$), see Fig. \ref{dissdoubtrans}.
As for the escape rate, the time evolution of the number of the
surviving particles is similar to the case without dissipation for the
short time, i.e., the escape rate is constant (the inset in
Fig. \ref{dissdoubtrans}(b)), while for the long time the escape rate
shows monotonic increase, see Fig. \ref{dissdoubtrans}. This is in
accordance with the general behavior in the doubly transient chaos
\cite{Motter2013,Chen2017}.
Note that this behavior is regularization dependent,
since $r_\text{min}$ becomes very close to zero near the peaks upon
introducing the dissipation, see Fig. \ref{dissdoubtpass}(b).

\section{Stability against the mass deformation}
\begin{figure}
 \centering
 \includegraphics[width=0.5\hsize]{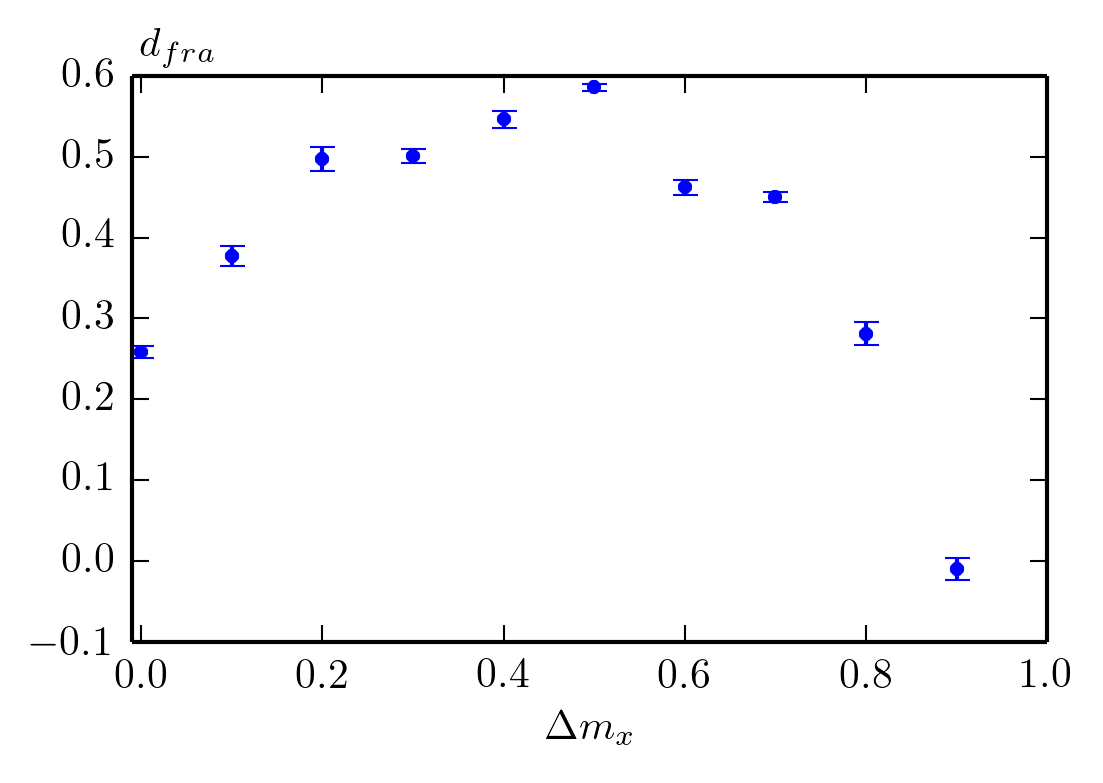}
 \caption{The fractal dimension against the mass deformation,
 $\Delta m_x$. $\Delta m_x=m_x-1$ and $\Delta m_x=0$ is the original
 parameter. By increasing $\Delta m_x$, the mass along $x$ direction
 becomes large and the system becomes more and more anisotropic.}
  \label{massdef}
\end{figure}
Since the rotational symmetry along $z$ direction leads to the reduction
of one degree of freedom, it is important to verify the stability
against the deformation which breaks the symmetry. We calculated the
variation of the fractal dimension under increasing the mass along $x$
direction, see Fig. \ref{massdef}. As we can see, the fractal dimension
remains positive against the finite amount of the perturbation, and then
becomes zero. Therefore, we expect our result is valid even in the
realistic situation where the rotational symmetry along $z$ direction is
broken.

%\bibliographystyle{apsrev4-1}
%\bibliography{reference}

 \newcommand{\noop}[1]{}